\documentclass[12pt]{article}
\usepackage{ulem}
\usepackage{times}
\usepackage{graphicx}
\usepackage{calc}
\usepackage{epsfig}
\usepackage{float}

\begin{document}

\parskip 18pt
\baselineskip 0.25in

\title{Type I and type II neuron models are selectively driven by
differential stimulus features}

\author{Germ\'an Mato and In\'es Samengo}

\date{Comisi\'on Nacional de Energ\'{\i}a At\'omica and Consejo de
Investigaciones Cient\'{\i}ficas y T\'ecnicas\\
Centro At\'omico Bariloche and Instituto Balseiro\\
8400 San Carlos de Bariloche, R. N., Argentina}

\maketitle

\vspace*{2cm}

{\center \Large \bf Abstract}\\ \vspace{0.1cm} \\
Neurons in the nervous system exhibit an outstanding variety of
morphological and physiological properties. However, close to
threshold, this remarkable richness may be grouped succinctly into
two basic types of excitability, often referred to as type I and
type II. The dynamical traits of these two neuron types have been
extensively characterized. It would be interesting, however, to
understand the information-processing consequences of their
dynamical properties. To that end, here we determine the differences
between the stimulus features inducing firing in type I and in type
II neurons. We work both with realistic conductance-based models and
minimal normal forms. We conclude that type I neurons fire in
response to scale-free depolarizing stimuli. Type II neurons,
instead, are most efficiently driven by input stimuli containing
both depolarizing and hyperpolarizing phases, with significant power
in the frequency band corresponding to the intrinsic frequencies of
the cell.

\newpage

\section{Introduction}

Several research lines have recently used reverse correlation
methods to determine the stimulus features that are most relevant in
shaping the probability to generate spikes of sensory neurons. Just
to mention a few examples, in the visual system, Fairhall,
Burlingame, Narasimhan, Harris, Puchalla and Berry (2006) revealed
multiple spatio-temporal receptive fields (STRF) driving salamander
retinal ganglion cells. In turn, Rust, Schwartz, Movshon, and
Simoncelli (2005) explored the STRF of macaque V1. In the
somatosensory system, Maravall, Petersen, Fairhall, Arabzadeh, and
Diamond (2007) employed covariance analysis to disclose the effect
of adaptation on the shift of coding properties in rat barrel
cortex.

Here, we are interested in exploring the way the relevant stimulus
features driving neuronal firing depend on the intrinsic dynamical
properties of the cell. To that end, we work with a time-dependent
stimulus representing the total input current arriving at the axon
hillock. In any real system, this input current may enter into the
cell in several ways. For example, it can pass through receptor
channels, activated by specific physical agents (light, sound,
temperature, etc.) relevant to a particular sensory modality.
Alternatively, it may be the integrated synaptic current entering a
central cell through its numerous dendrites, or through an
intracellular electrode. In any case, we shall assume that our
time-dependent stimulus $s(t)$ represents a ionic current that,
after propagating all along the spatial extension of the neuron,
arrives into the axon hillock, where the decision to fire or not to
fire is taken. Hence, we shall only be dealing with the temporal -
and not spatial - properties of the input current.

Our aim is to understand the relation between the stimulus
attributes that most strongly affect the firing probability and the
intrinsic dynamical properties of the neurons. This line of research
was initiated with the study of the relevant stimulus features
driving integrate-and-fire model neurons (Ag\"uera y Arcas \&
Fairhall, 2003) and in Hodgkin-Huxley cells (Ag\"uera y Arcas,
Fairhall, \& Bialek 2003). Later on, Hong, Ag\"uera y Arcas and
Fairhall (2007) explored a broader class of neuron models,
determining the effect of several dynamical features on the relevant
stimulus dimensions. Here, we extend those analysis, searching for
unifying properties that characterize the way in which the type of
bifurcation at firing onset affects stimulus selectivity and
information transmission. We therefore compare the relevant stimulus
features driving two broad classes of neuron models, namely, type I
and type II dynamics. The distinction between type I and type II
excitability was first introduced by Hodgkin (1948), when studying
the dependence of the firing rate of a neuron with the injected
current. Later, more detailed classifications of neuronal
excitability were introduced (Ermentrout 1996, Izhikevich 2007), in
terms of the bifurcation type at firing onset. In all cases, type I
cells undergo a saddle-node bifurcation on the invariant circle, at
threshold. Type II neurons, instead, may correspond to 3 different
bifurcations, namely, a subcritical Hopf bifurcation, a
supercritical Hopf bifurcation, or a saddle node bifurcation outside
the invariant circle. Most of type II conductance-based
Hodgkin-Huxley type neuron models, however, undergo a subcritical
Hopf bifurcation. Therefore, here, when exploring type II neuron
models, we shall focus on those exhibiting a subcritical-Hopf
bifurcation.

\begin{figure}[htdf]
\centerline{\includegraphics[keepaspectratio=true, clip = true,
scale = 1.0, angle = 0]{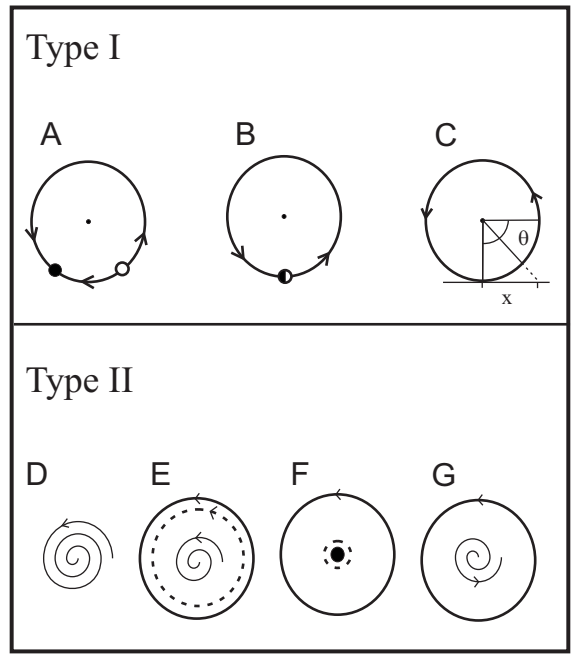}} \caption{\label{fig1} Schematic
representation of the bifurcation underlying the onset of firing in
type I (A, B, C) and type II (D, E, F, G) neuronal models. The input
current is increased from left to right, and in both cases, the
system passes from a single fixed point (A and D) describing the
subthreshold rest state, to a firing limit cycle (C and G).}
\end{figure}

When type I neurons are stimulated by a constant input current, the
onset of firing is brought about by a saddle node bifurcation
(Izhikevich, 2007). In the upper panel of Fig.~\ref{fig1}, a
schematic representation of the topology of a type I bifurcations is
depicted. For subthreshold input currents ({\em A}), the system
contains two orbits forming a closed figure, connecting two fixed
points: one of them is stable, and the other is unstable. As the
injected current $I$ increases, the two fixed points come nearer to
each other, so that when $I$ reaches a critical value $I_{\rm c}$,
the two points coalesce into a single one ({\em B}). Thereafter,
both equilibrium points disappear, and the system moves in a
periodic orbit ({\em C}).

The topological properties of the subcritical Hopf bifurcation (also
called {\sl inverted} Hopf bifurcation) characteristic of many type
II models are shown in the lower panel of Fig.~\ref{fig1}. For
subthreshold input currents, there is only a single stable spiral
fixed point ({\em D}). As the current increases, there is a region
far away from the fixed point where the radial velocity diminishes.
At a certain critical current $I_{\rm g}$, two closed orbits appear
through a global bifurcation, the outermost stable, the inner one
unstable ({\em E}). As the current is increased further, the
unstable limit cycle shrinks, approaching the spiral fixed point
({\em F}). When $I$ reaches a second critical value $I_{\rm c}$, the
unstable orbit coalesces onto the fixed point. If $I$ grows beyond
$I_{\rm c}$, the fixed point is turned into an unstable spiral node,
and the only stable attractor of the system is the distant limit
cycle ({\em G}).

These two types of cells differ from each other in the bifurcation
underlying the onset of firing. This comprises a clear topological
difference, that endows each type with specific dynamical
properties. It would be interesting, however, to be able to identify
the functional consequences of these dynamical differences. More
precisely, we would like to determine in which way the two neuron
types differ, regarding their information-processing properties.
Specifically, what kind of time-dependent stimulus is needed to
induce spiking in a type I neuron, and how does this stimulus differ
from the one needed to excite a type II neuron? To answer this
question, we explore the relevant stimulus features shaping the
firing probability of type I and type II neurons, by means of
covariance analysis. In the first place, we work with
conductance-based models, capturing the biophysically relevant
processes. We then compare the results obtained for these realistic
models with those derived from reduced neural models, which
minimally describe the essential dynamical features of both neuronal
types. We conclude that close to threshold, the relevant stimulus
features driving a given cell are determined by the type of
bifurcation.

\section{The models and their phase-resetting curves}

\label{models}

In the case of type I neurons, we use the model neurons introduced
by Wang and Buzsaki (1996) to describe hippocampal interneurons.
This model (hereafter called ``WB'') when stimulated with a
supra-threshold constant input current settles into a limit cycle,
embedded in its 3-dimensional phase space. As the input current is
lowered, the firing frequency tends to zero, and the system displays
a saddle-node bifurcation into its resting state. Therefore, the WB
model can be classified as type I. The detailed equations are
displayed in Appendix A. The second model is the original
Hodgkin-Huxley (HH) neuron (see the equations in Appendix B). This
is a 4-dimensional system which, at threshold, undergoes a
subcritical Hopf bifurcation, and is therefore classified as type
II.

Figure ~\ref{fig1} shows the qualitative behavior of type I and type
II neural models upon constant stimulation. We see that both systems
oscillate periodically if the input is above the critical current
$I_{\rm c}$. The behavior of these systems when weakly perturbed by
rapid current injections is characterized by the phase-response
function $Z(t)$. This function describes how the phase of a spike is
advanced or delayed, depending on the time $t$ at which the system
is perturbed (Kuramoto, 2003). More precisely, $Z(t)$ is defined as
the ratio between the change of the phase and the size of the
perturbation in the limit of small perturbations. In principle the
perturbation can be applied along any of the variables of the
dynamical system, giving rise to a multi-dimensional phase-response
function. In this paper, however, perturbations are only
instantiated as input currents, thereby only affecting a single
direction in phase space: the one parallel to the membrane potential
axis $V$. Hence, here, $Z(t)$ represents the magnitude of a
phase-response function always pointing in the direction of $V$ (or
$-V$, if $Z(t)$ is negative). By convention, we stipulate that the
time of spike generation (defined as the point where the voltage
crosses the value 0 mV from below) corresponds to $t = 0$. The
function $Z(t)$ has the same period $T$ as the firing cycle of the
neuron which, in turn, depends on the size of the constant input
current.

\begin{figure}[htdf]
\centerline{\includegraphics[keepaspectratio=true, clip = true,
scale = 1.0, angle = 0]{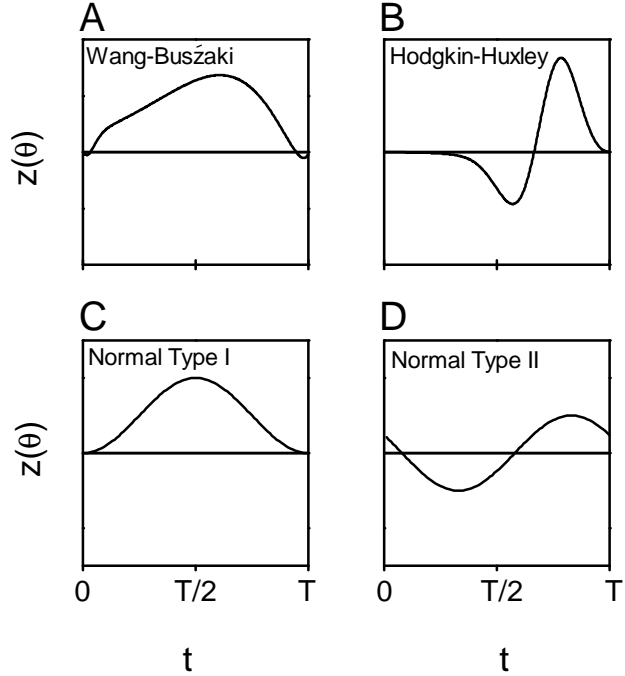}} \caption{\label{fig2} Type I
(left panels) and type II (right panels) phase response curves. The
realistic models (WB and HH) are shown in the upper row, and the
simplified normal forms are depicted in the bottom row.}
\end{figure}

In panels {\em A} and {\em B} of Fig.~\ref{fig2} the phase-response
curve of the WB and HH models is depicted. The most salient
difference between the two models is that $Z(t)$ is (almost) always
positive for the WB model, whereas it exhibits both positive and
negative regions in the HH case (Hansel, Mato \& Meunier, 1995).
Hence, a small depolarizing perturbation always results in
anticipated firing in the WB model, whereas it may either advance or
delay spiking in the HH model, depending on when exactly the
perturbation is delivered. Both the WB and HH models show maximal
sensitivity to the external perturbation far away from the spiking
events. In addition, the HH model shows more evidently that right
after the spike ($t \approx 0$), the system is rather unresponsive
to incoming perturbations, due to refractoriness.

In the upper panels of Fig.~\ref{fig2}, the DC component of the
input current was fixed to a different value in each model, so as to
obtain a firing rate of 62.5 Hz in both cases (that is, $T = 16$
ms). When the DC input current is lowered, the HH model eventually
enters in its bistable region at a finite firing frequency. The WB
model, instead, can be driven with arbitrarily low firing rates.
Actually, as the firing rate of the WB model is diminished, the
small region of negative $Z(t)$ that is observed for $t \approx T$
and $t \approx 0$ disappears, leaving a purely positive
phase-resetting curve. In fact, only near threshold do type I
neurons display a purely positive phase-resetting; if the input
current increases, the curve begins to resemble that of type II
neurons.


Near threshold, hence, the HH neuron shows a phase-resetting curve
that is qualitatively different from that of the WB neuron. The
question now arises whether this difference is specific to these two
neuron models, or whether it is always found when comparing a type I
with a type II cell. We therefore analyze the phase-resetting curves
of the reduced models. To that end, the input current is written as
\begin{equation}
I = I_{\rm c} + \epsilon^2 i.
\end{equation}
Ermentrout \& Kopell (1986) and Ermentrout (1996) have shown that
the dynamics of any type I neuron can be reduced to a 1-dimensional
equation for a variable $x$ that represents the projection of the
state of the system on the direction of phase space that loses
stability at threshold. After a non-linear transformation $x =
\tan(\theta/2)$, and rescaling the temporal variable this dynamics
can be written as
\begin{equation}
\frac{{\rm d}\theta}{{\rm d}t} = \left[1 - \cos(\theta)\right] +
\alpha \left[1 + \cos\theta \right], \label{northet}
\end{equation}
where $\alpha = \eta/2 \epsilon q$ plays the role of a bifurcation
parameter. It is defined in terms of $q$ (depending solely on the
dynamical properties of the original model) and $\eta$ (proportional
to the magnitude of the perturbation $i$). Both parameters may be
derived from the equations of the full-blown system. Spike
generation is associated to $\theta = \pm \pi$ (see Fig.
\ref{fig1}). Notice that Eq. (\ref{northet}) is valid for any neural
model sufficiently close to a saddle-node bifurcation. Hence,
irrespective of the diversity of dynamical richness, near firing
onset the whole collection of type I conductance-based models is
topologically equivalent to a system controlled by just a single
parameter $\alpha$.

For 1-dimensional systems, the phase response curve can be evaluated
analytically (Hansel, Mato \& Meunier, 1995; Kuramoto, 2003). For this case it is
found that
\begin{equation}
Z(t) \propto \sin^2\left(\pi t / T \right), \label{prc1}
\end{equation}
as depicted in Fig.~\ref{fig2}{\sl C}. Notice that, as in the WB
model, $Z(t)$ is always positive, with maximum sensitivity near $t =
T/2$. Therefore, up to some positive scaling constant, the response
function of type I models near the bifurcation is universal. In what
follows, all numerical integrations of the reduced (or {\sl normal})
type I system are carried out with Eq.~(\ref{northet}).


Why is the phase response function always positive for type I
neurons, and why does it approach zero in the vicinity of spiking
times? Near the bifurcation, type I neurons spend most of their time
in the neighborhood of $\theta = 0$ or $x = 0$ (that is, far away
from the spiking region $\theta = \pi$ or $x \to \pm \infty$). The
slow dynamics near $\theta = 0$ is a footprint of the two fixed
points that appear through a saddle-node bifurcation, when the input
current is lowered below the critical value. As a consequence, in
the slightly supra-threshold regime, the system spends a large
fraction of each period around $x = 0 = \theta$. Hence, almost all
the perturbations find the system in this region, where positive
perturbations advance the phase of the neuron. Therefore, for almost
all $t$, external perturbations result in anticipated spike
generation and thereby, in a positive phase-response curve. In
addition, in the vicinity of $\theta = \pm \pi$, the system sets
into the rapid acceleration associated with spike generation. In
this region, the dynamics is dominated by the catalytic opening of
voltage-dependent conductances, and not by the detailed temporal
properties of the perturbing current. Hence, $Z(t) \to 0$, when $t
\approx T$.

We now turn to the reduced model of the HH neuron. In this case, the
onset of firing is governed by a subcritical Hopf bifurcation. Only
two of the four eigenvalues lose stability at threshold, meaning
that the bifurcation takes place in two dimensions. Hence, the
reduced model is 2-dimensional, and similarly to Brown et al.
(2004), we choose to analyze it in polar coordinates
\begin{eqnarray}
\frac{{\rm d}r}{{\rm d}t} &=& \alpha r + c r^3 + f r^5 \label{ho1} \\
\frac{{\rm d}\phi}{{\rm d}t} &=& 2\pi( \beta + d r^2 + g r^4)
\label{ho2}
\end{eqnarray}
where, for subcritical bifurcations, $c > 0$ , $f < 0$. Notice that
the reduced system described by Eqs.~(\ref{ho1}) and (\ref{ho2})
represents a combination of two bifurcations: a non-local
saddle-node bifurcation of limit cycles far away from the fixed
point (represented in Fig.~\ref{fig1}{\em E}), and a local
subcritical Hopf bifurcation (Fig.~\ref{fig1}{\em F}). This means
that the parameters defining the spatially extended system
(\ref{ho1}) and (\ref{ho2}) cannot be obtained by a local reduction
of the full-blown model. Therefore, they have to be understood as an
approximate representation of the original system, with its same
topological properties.

The bifurcation parameter $\alpha$ is proportional to $I - I_{\rm
c}$. When $\alpha > 0$ the fixed point $r = 0$ is unstable, and all
trajectories tend to a limit cycle located at $r = [-c(1 + \sqrt{1 -
4\alpha f / c^2})/2 f]^{1/2}$, corresponding to the regular firing
trajectory. If $\alpha$ is decreased such that $c^2/4 f < \alpha <
0$, then $r = 0$ becomes a stable fixed point, and it coexists with
a stable limit cycle at $r = [-c(1 + \sqrt{1 - 4 \alpha f / c^2}) /
2 f]^{1/2}$. The two attractors are separated by an unstable limit
cycle located at $r = [-c(1 - \sqrt{1 - 4 \alpha f})/2 c^2]^{1/2}$.
If $\alpha$ is decreased even further, such that $\alpha < c^2/4 f$,
a single stable manifold remains: a fixed point at $r = 0$,
representing the subthreshold resting state.

The stable limit cycle of the firing trajectory has a minimal radius
of $\sqrt{-c/2f}$, when $\alpha = c^2 / 4 f$. Therefore, spike
generation will be associated to the moment where the system crosses
the border $\theta = \pi$ with a radius $r \ge \sqrt{-c/2f}$.

As before, the phase resetting curves can be evaluated analytically
(Brown, Moehlis, \& Holmes, 2004)
\begin{equation}
Z(t) \propto \sin\left[2\pi (t-t_0) / T\right], \label{prc2}
\end{equation}
where $t_0$ and the proportionality constant can be evaluated in
terms of the parameters of the original HH model.
Fig.~\ref{fig2}{\sl D} depicts the phase response curve of
Eq.~(\ref{prc2}). As observed in the full HH model, external
perturbations may either advance or delay spiking, depending on when
in the cycle they are delivered. By comparing the phase-response
curves in Fig.~\ref{fig2}{\sl A} and {\sl C}, and those in {\sl B}
and {\sl D}, we conclude that the qualitative features of the
biophysically realistic models are captured by the reduced models.


In this work, we are interested in relating phenomenological
description of the input/output mapping carried out by a given cell
with the dynamical properties of the cell. In this context, it is
interesting to point out that type I and type II neuron models have
qualitatively different phase-response curves. This property
suggests that the two neural types are selective to different
stimulus features. This hypothesis was confirmed by Ermentrout,
Gal\'an, \& Urban (2007), where they indicate that in
quasi-periodically oscillating neurons that are weakly perturbed by
input noise, the spike-triggered average is proportional to the
derivative of the phase-response curve. Hence, when a neuron is
stably circling around its firing limit cycle, the shape of the
phase-response curve is highly informative of the stimulus features
that induce spiking. In this paper, however, we are interested in
analyzing the behavior of neuron models in the vicinity of firing
onset. Therefore, the results derived in the supra-threshold regime
are not necessarily applicable. In fact, Ag\"uera y Arcas, Fairhall,
\& Bialek (2003) have shown a spike-triggered average in
Hodgkin-Huxley neurons in the excitable regime that strongly
resembles the phase-response curve itself (see Fig. \ref{fig2}), and
not its derivative. Both studies, hence, indicate that the
phase-response curve contains information about the relevant
stimulus features inducing spiking. The exact relationship between
the two quantities, however, seems to depend critically on the mean
level of depolarization. In this paper, we disclose the optimal
stimulus features driving a cell that is initially near its resting
state, and that only occasionally generates action potentials. We
therefore work with highly variable spike trains, as opposed to
Ermentrout, Gal\'an \& Urban (2007).


\section{Covariance analysis and the extraction of relevant stimulus features}

In order to explore the stimuli that are most relevant in shaping
the probability of spiking, we use spike-triggered covariance
techniques (Bialek \& De Ruyter von Steveninck, 2003; Paninski,
2003; Schwartz, Pillow, Rust \& Simoncelli, 2006).
Our purpose is to relate the results obtained from this standard
statistical approach (which essentially treats the cell as a black
box operating as an input/output device) with the internal (that is,
dynamical) properties of the neurons.

If $P[{\rm spike \ at \ } t_{\rm 0} | s(t)]$ is the probability to
generate a spike at time $t_{\rm 0}$ conditional to a time-dependent
stimulus $s(t)$, we assume that $P$ only depends on the stimulus
$s(t)$ through a few relevant {\em features} $f^1(t - t_{\rm 0}),
f^2(t - t_{\rm 0}), ..., f^k(t - t_{\rm 0})$. The stimulus and the
relevant features are continuous functions of time. For
computational purposes, however, we represent them as vectors ${\bf
s}$ and ${\bf f}^i$ of $N$ components, where each component $s_j =
s(j \delta t)$ and $f^i_j = f^i(j \delta t)$ is the value of the
stimulus evaluated at discrete intervals $\delta t$. If $\delta t$
is small compared with the relevant time scales of the models this
will be a good approximation.

The relevant features ${\bf f}^1...{\bf f}^k$ lie in the space
spanned by those eigenvectors of the matrix
\begin{equation}
M = C_{\rm prior}^{-1} C_{\rm spikes}, \label{matrix}
\end{equation}
whose eigenvalues are significantly different from unity  (Schwartz,
Pillow, Rust \& Simoncelli, 2006). Here, $C_{\rm spikes}$ is the $N
\times N$ spike-triggered covariance matrix
\begin{equation}
(C_{\rm spikes})_{ij} = \frac{1}{N_{\rm spikes}}\sum_{t_0} s(t_i+
t_0)s(t_j + t_0) - s^0(t_i)s^0(t_j), \label{cspikes}
\end{equation}
where the sum is taken over all the spiking times $t_0$, and
$s^0(t)$ is the spike-triggered average (STA)
\begin{equation}
s^0(t) = \frac{1}{N_{\rm spikes}} \sum_{t_0} s(t + t_0).
\label{esta}
\end{equation}
Similarly, $C_{\rm prior}$ (also with dimension $N\times N$) is the
prior covariance matrix
\begin{equation}
(C_{\rm prior})_{ij} = \overline{s(t_i + t)s(t_j + t)} -
(\bar{s})^2, \label{cprior}
\end{equation}
where the horizontal line represents a temporal average on the
variable $t$.

The eigenvalues of $M$ that are larger than 1 are associated to
directions in stimulus space where the stimulus segments associated
to spike generation have an increased variance, as compared to the
raw collection of stimulus segments. More precisely, the eigenvalue
itself provides a measure of the ratio of variances of the two
ensembles. Correspondingly, those eigenvalues that lie significantly
below unity are associated to stimulus directions of decreased
variance. That is, an eigenvalue that is noticeably smaller than
unity indicates that there is a certain feature for which the ratio
of the variance of the spike-triggering stimuli and the variance of
the raw stimuli is significantly small.

In order to perform a covariance analysis of a specific neuronal
model, we simulate the cell with an input current
\begin{equation}
I(t) = I_{\rm 0} + \sigma \xi(t). \label{curr}
\end{equation}
The DC term $I_0$ lies slightly below threshold, so that the cell is
not able to generate spikes in the absence of the Gaussian noise
term $\xi(t)$. The firing rate is regulated by adjusting $\sigma$.
The noise $\xi(t)$ is such that $\langle \xi(t) \rangle = 0$ and
$\langle \xi(t) \xi(t')\rangle = \tau \exp(-|t-t'|/\tau) / 2$. In
what follows, we use $\tau = 0.1$ or $0.2$ ms, that is much faster
than the characteristic time constants of the models, and much
slower than the numerical integration time 0.01 ms.

The input current given by Eq.~(\ref{curr}) is injected into the
realistic models (WB and HH) as an additive term in the equations
governing the temporal evolution of the voltage variable (see
Eqs.~(\ref{wb1}) and (\ref{hh1}), in the appendices). For the type I
model, $\eta$ is proportional to the injected current $I(t)$ of
Eq.~(\ref{curr}). Hence, in Eq.~(\ref{northet}), $\eta$ is replaced
by $I(t)$, with $I_{\rm 0}$ slightly below zero. Spike generation is
identified with $\theta = \pi$. In the case of type II neurons,
$I(t)$ should be parallel to the voltage variable. However, the
2-dimensional system of Eqs. (\ref{ho1}) and (\ref{ho2}) does not
need to contain the voltage axis of the original HH system, since
none of the two eigenvectors associated to the two eigenvalues that
lose stability at threshold is necessarily parallel to the voltage
axis. However, the voltage axis is not orthogonal to this
2-dimensional system. Hence, if an input current drives the original
HH system, that current has a non-zero projection onto the reduced
2-dimensional system. The reduced system should therefore be excited
in the direction in which the original voltage axis projects onto
the subspace spanned by the eigenvectors associated to the two
eigenvalues that loose stability. In our simulations, the input
current (\ref{curr}) was injected making a 45$^\circ$ angle with the
x-axis. We have verified, however, that our results do not depend
critically on the value of this angle.

\section{Relevant stimulus features driving type I neurons}
\label{type1}

\begin{figure}[htdf]
\centerline{\includegraphics[keepaspectratio=true, clip = true,
scale = 1.0, angle = 0]{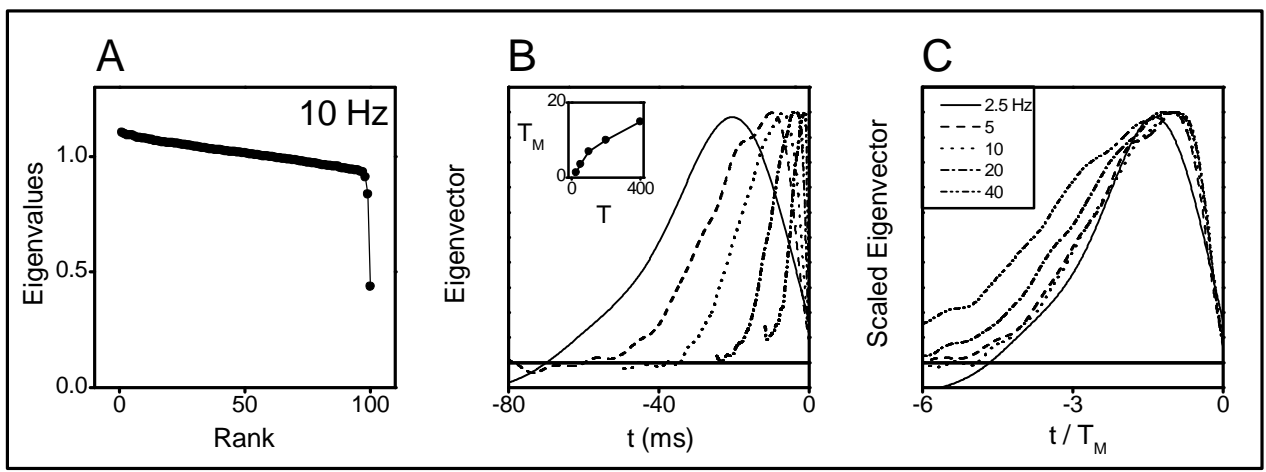}} \caption{\label{fig3} Covariance
analysis of the WB model. In all cases, $I_0 = 0$, $\tau = 0.2$ ms.
{\em A}: Eigenvalue spectrum obtained for 10 Hz ($\sigma$ =
15.5$\mu$A/cm$^2$ms$^{1/2}$, CV = 0.74, 50420 spikes). The
qualitative aspect of the spectrum of eigenvalues remains the same,
for all firing rates shown in {\em B} and {\em C}. {\em B}:
Eigenvectors corresponding to the smallest eigenvalue. Different
curves are obtained by setting the noise $\sigma$ to different
values, and thereby, by varying the firing rate: 2.5 Hz ($\sigma$ =
8.1 $\mu$A/cm$^2$ms$^{1/2}$, CV = 0.87, 55099 spikes), 5 Hz
($\sigma$ = 10.3 $\mu$A/cm$^2$ms$^{1/2}$, CV = 0.81, 50956 spikes),
10 Hz ($\sigma$ = 15.5 $\mu$A/cm$^2$ms$^{1/2}$, CV = 0.87, 55099
spikes), 20 Hz ($\sigma = 25.5 \mu$A/cm$^2$ms$^{1/2}$, CV = 0.74,
50790 spikes), 40 Hz ($\sigma$ = 69.0 $\mu$A/cm$^2$ms$^{1/2}$, CV =
0.81, 49110 spikes). As the firing rate grows, the eigenvector
becomes increasingly narrow. Line conventions shown in {\em C}. {\em
Inset}: Time $T_{\rm M}$ at which each eigenvector reaches its
maximum value, as a function of the mean inter-spike interval $T$
(the inverse of the firing rate). {\em C}: Scaled eigenvectors, for
different firing rates. Hence, at firing onset the WB model is
sensitive to a universal depolarizing stimulus, whose temporal scale
depends solely on the firing rate of the cell.}
\end{figure}

The results of a covariance analysis of the WB model are depicted in
figure \ref{fig3}.
In panel {\em A}, the spectrum of eigenvalues of a 10 Hz firing
neuron is shown. The majority of the eigenvalues cluster around
unity. There is, however, a single eigenvalue that is notoriously
smaller than all the others. This spectrum remains essentially
unchanged, as the firing rate of the neuron is varied between 2.5
and 40 Hz. In {\em B}, the eigenvector corresponding to the smallest
eigenvalue shown in {\em A} is depicted in a dotted line. The other
lines show how this eigenvector changes, as the firing rate of the
cell is modified. This is accomplished by fixing the noise $\sigma$
to different values. Hence, the different curves in {\em B}
correspond to the single relevant eigenvector obtained for different
firing rates (see the line conventions in {\em C}).

For all firing rates, we see that there is a single eigenvalue that
is significantly smaller than unity. This means that the relevant
eigenvector corresponds to a stimulus direction with diminished
variance. In all cases, the relevant eigenvector shows an unimodal
curve, that is either always positive, or always negative (recall
that the sign of an eigenvector is not determined by the covariance
analysis). In Fig.~\ref{fig3} we have chosen to represent the
eigenvectors as positive (and not negative) stimulus segments,
because in this way they coincide with the STA (data not shown).
This allows us to conclude that the WB neuron model fires in
response to depolarizing stimuli. This result is in agreement with
the phase-resetting curves obtained for the WB model (see
Fig.~\ref{fig2}{\sl A}). Hence, even though phase-resetting curves
(necessarily calculated with stimuli of supra-threshold mean) and
our covariance analysis (carried out with stimuli of sub-threshold
mean) correspond to two different firing regimes, there is a
qualitative parallelism between them: in type I neurons near their
firing onset, depolarizing input currents always favor spike
generation.

In Fig~\ref{fig3}{\em B} we see that as the firing rate increases,
the relevant stimulus feature reduces the span of its temporal
domain. By naked eye, it seems that the main effect of changing the
firing rate is a temporal rescaling of the eigenvector. In order to
check this hypothesis, for each firing rate we determine the time
$T_{\rm M}$ at which the relevant eigenvector reaches its peak
value. In the inset of panel {\em B}, $T_{\rm M}$ is depicted as a
function of the mean inter-spike interval $T$ (the inverse of the
firing rate). Next, we re-scale each eigenvector, by plotting it as
a function of $t / T_M$, as shown in {\em C}. There we see that
throughout a 16-fold increase in the firing rate, the shape of the
relevant eigenvector remains essentially unchanged, apart from a
temporal scaling. This implies that near threshold, spike generation
in the WB model is governed by a single relevant stimulus feature of
universal shape.

\begin{figure}[htdf]
\centerline{\includegraphics[keepaspectratio=true, clip = true,
scale = 1.0, angle = 0]{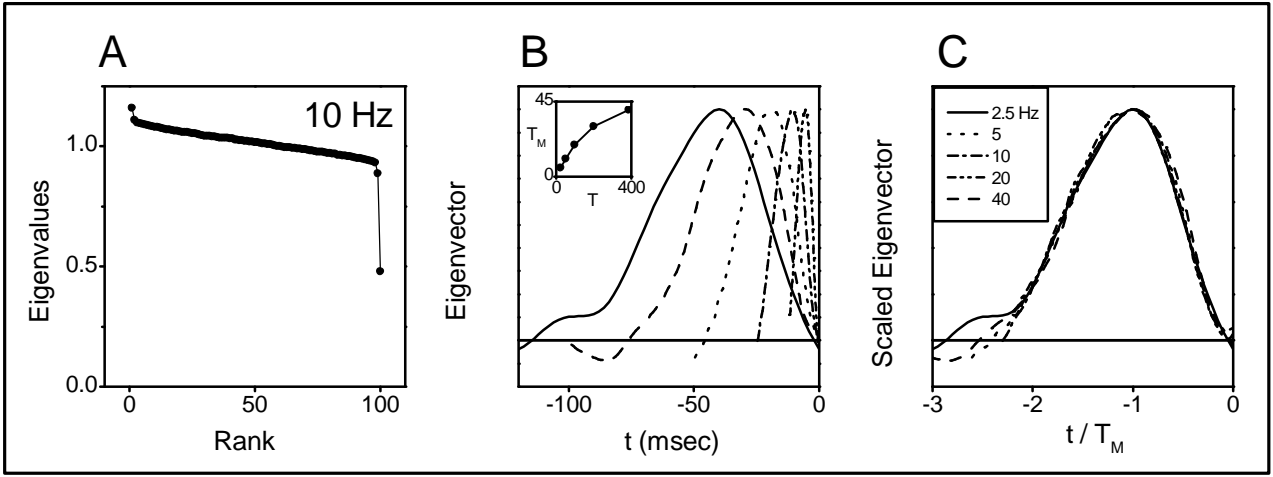}} \caption{\label{fig4} Covariance
analysis of the normal form of type I neurons. Spike generation is
identified as $\theta$ crossing the angle $\pi$. In all cases,
$I_{\rm DC} = -0.01 \mu$A/cm$^2$, and $\tau = 0.2$ ms. {\em A}:
Eigenvalue spectrum obtained for 10 Hz ($\sigma$ = 0.0046
$\mu$A/cm$^2$ms$^{1/2}$, CV = 0.64, 50607 spikes). {\em B}:
Eigenvectors corresponding to the smallest eigenvalue. Different
curves correspond to different firing rates: 2.5 Hz ($\sigma$ =
0.002935 $\mu$A/cm$^2$ms$^{1/2}$, CV = 0.81, 51718 spikes), 5 Hz
($\sigma$ = 0.00353 $\mu$A/cm$^2$ms$^{1/2}$, CV = 0.73, 49855
spikes), 10 Hz ($\sigma$ = 0.0046 $\mu$A/cm$^2$ms$^{1/2}$, CV =
0.64, 50607 spikes), 20 Hz ($\sigma$ = 0.00715
$\mu$A/cm$^2$ms$^{1/2}$, CV = 0.60, 49962 spikes), 40 Hz ($\sigma$ =
0.01421 $\mu$A/cm$^2$ms$^{1/2}$, CV = 0.57, 50296 spikes). {\em
Inset}: Time $T_{\rm M}$ at which each eigenvector reaches its
maximum value, as a function of the mean inter-spike interval $T$
(the inverse of the firing rate). {\em C}: Scaled eigenvectors, for
different firing rates. The normal type I model, hence, shows the
same qualitative behavior as the conductance-based WB model.}
\end{figure}

How general is this universality? Following Ermentrout \& Kopell
(1986) and Ermentrout (1996), in the previous section we pointed out
that as an arbitrary type I neuron approaches threshold, its
dynamical equations can be reduced to the normal form Eq.
(\ref{northet}). Recall that the single parameter appearing in
Eq.~(\ref{northet}) is proportional to the (scaled) perturbative
input current $i$. Though this reduction is only valid under
constant stimulation, the absence of typical time-scales in the
normal form of type I neurons suggests that perhaps, the universal
eigenvector shown in Fig.~\ref{fig3} might be a general property of
type I models near firing onset. In order to check this hypothesis,
in Fig.~\ref{fig4}
we depict the results of performing a covariance analysis of the
normal-form dynamical model of Eq.~(\ref{northet}). In {\em A}, the
eigenvalue spectrum obtained for 10 Hz firing rate is shown. As in
the WB model, the most salient eigenvalue is significantly smaller
than unity. The largest eigenvalue also seems to depart from unity.
We have checked, however, that as one gets closer to threshold (as
$I_{\rm DC} \to 0^-$), this eigenvalue approaches unity.

Just as it happened with the WB model, near threshold the spectrum
of eigenvalues remains essentially unchanged, as the firing rate is
varied: in all cases, there is a single eigenvalue significantly
smaller than unity. In {\em B}, the eigenvector corresponding to
this eigenvalue is depicted, for different values of the firing
rate. The same qualitative behavior seen in Fig.~\ref{fig3} is
obtained: as the firing rate grows, the depolarizing fluctuation in
the eigenvector becomes increasingly time-compressed. In the inset
of panel {\em B}, we show the time $T_{\rm M}$ at which the
eigenvector reaches its maximum value, as a function of the mean
period $T$ (the inverse of the mean firing rate). In {\em C}, the
scaled eigenvectors may be seen to fit a fairly universal shape, in
spite of the 16-fold variation in the firing rates. Hence, we
conclude that the universal behavior observed in the WB model is
actually a general property of type I neurons, near threshold.

\section{Relevant stimulus features driving type II neurons}
\label{type2}

\begin{figure}[htdf]
\centerline{\includegraphics[keepaspectratio=true, clip = true,
scale = 1.0, angle = 0]{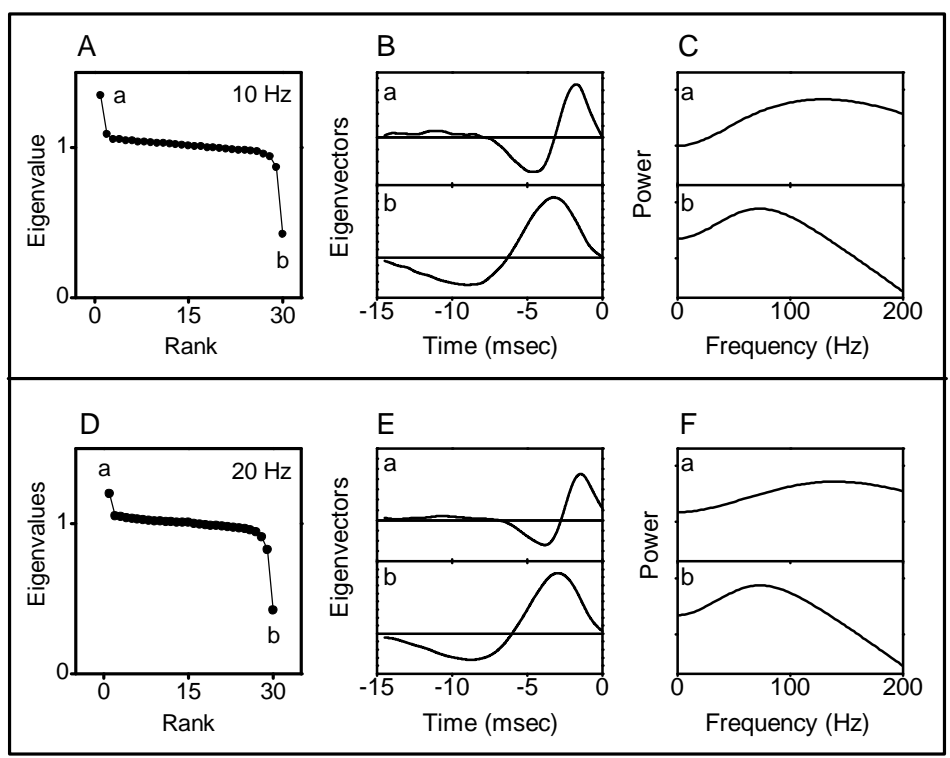}} \caption{\label{fig5} Covariance
analysis of the HH model neuron for two values of the firing rate,
for $\tau = 0.2$ ms. {\sl Upper panels}: 10 Hz firing rate ($\sigma=$17.2
$\mu$A/cm$^2$ms$^{1/2}$, CV = 0.85, 51165 spikes). {\sl Lower
panels}: 20 Hz firing rate ($\sigma=$22.15 $\mu$A/cm$^2$ms$^{1/2}$, CV =
0.69, 51490 spikes). {\em A} and {\em D}: spectra of eigenvalues. In
each case, two outliers are singled out, labeled as {\em a} and {\em
b}. {\em B} and {\em E}: Time-domain representation of the
eigenvectors corresponding to the eigenvalues {\em a} and {\em b}.
{\em C} and {\em F}: Frequency-domain representation of the
eigenvectors {\em a} and {\em b}.}
\end{figure}

In the previous section we saw that type I neurons fire in response
to depolarizing stimulus fluctuations, and that those fluctuations
lack an intrinsic temporal scale: the faster the stimulus, the
higher the firing rate. Here, we explore whether that is also the
case for type II neurons. In Fig.~\ref{fig5} we show the results
of carrying out a covariance analysis with the HH model neuron, for
two values of the firing rate: 10 Hz (upper panels), and 20 Hz
(lower panels). Each spectrum of eigenvalues ({\em A} and {\em D})
exhibits two outliers, labeled {\em a} and {\em b} in the figure.
The corresponding eigenvectors are depicted in {\em B} and {\em E}
(in the time domain) and in {\em C} and {\em F} (in the frequency
domain). In all cases, the eigenvectors exhibit both positive and
negative phases. This result is in agreement with the
phase-resetting curves obtained for the supra-threshold HH model
(Fig.~\ref{fig2}{\em B}), suggesting that the relevant subspace
comprises characteristic frequencies. For both firing rates, we
observe that each eigenvector contains a dominant frequency:
eigenvector {\em a} peaks at 125 Hz, whereas {\em b} reaches its
maximum at 62 Hz. As may be easily judged by comparing panels {\em
B} and {\em E}, the temporal domain of the eigenvectors does not
seem to contract, as the firing rate is doubled. Accordingly, the
principal frequency of each eigenvector (compare {\em C} and {\em
F}) remains unaltered. This means that in the HH model, each
eigenvector has a characteristic temporal pattern that is not
determined by the firing rate of the cell. A natural question,
hence, arises: What governs these temporal patterns and their
characteristic frequencies?

To answer this question, we apply covariance-analysis techniques to
the reduced model described by Eqs.~(\ref{ho1}) and (\ref{ho2}). In
the first place, we choose $c = 1$/ms and $f = -1$/ms. More
importantly, we fix $\dot{\theta} = \rm{cnst} = 2 \pi \beta$ (that
is, we make $g = d = 0$).
\begin{figure}[htdf]
\centerline{\includegraphics[keepaspectratio=true, clip = true,
scale = 1.0, angle = 0]{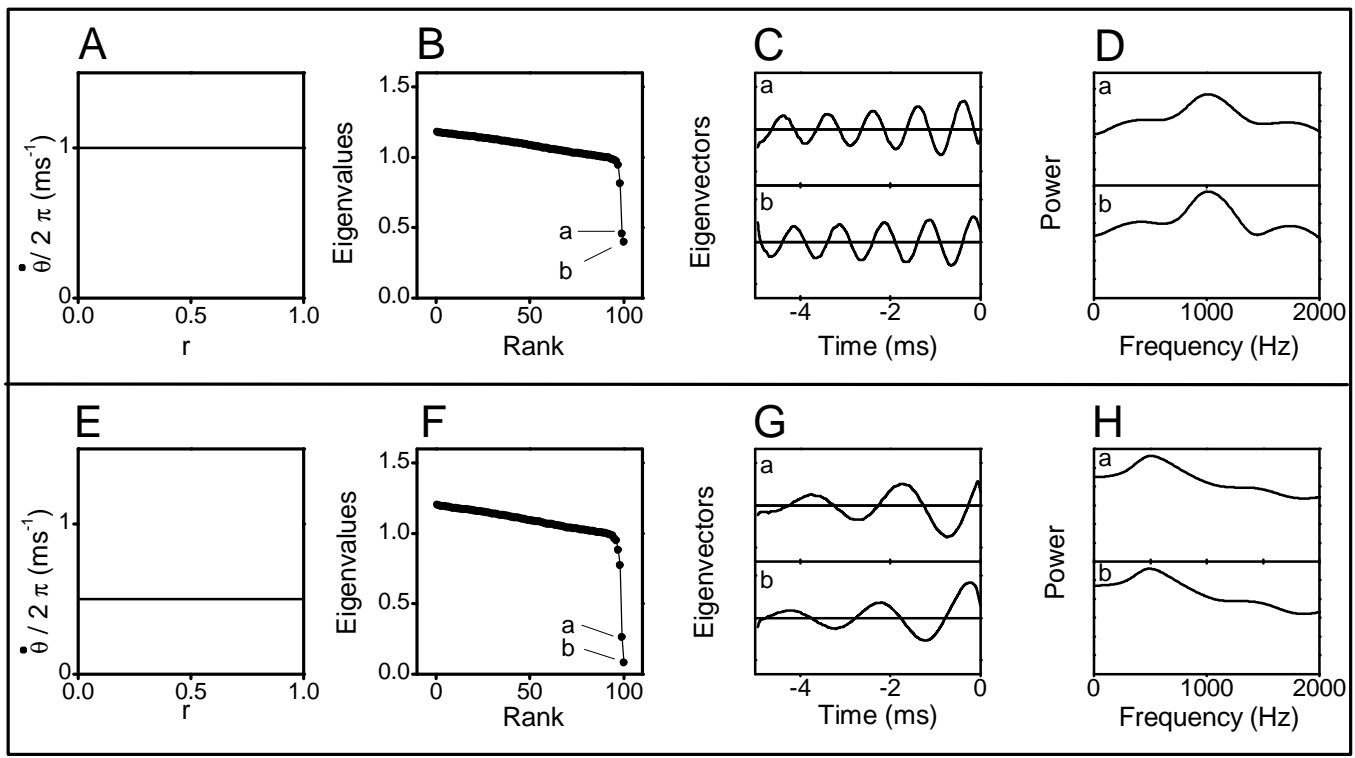}} \caption{\label{fig6} Covariance
analysis of a normal-form type II neuron model in which the angular
velocity $\dot{\theta}$ is independent of the radial coordinate. The
bifurcation parameter $\alpha$ is set to -0.5 / ms, and $\tau = 0.1$
ms. The noise $\sigma$ is chosen so as to obtain a firing rate of
200 Hz. Upper panels: $\beta = 1/$ ms, $\sigma = 0.037$ /
ms$^{1/2}$, CV = 2.3, 48246 spikes. Lower panels: $\beta = 0.5$ ms,
$\sigma = $ 0.0798 / ms$^{1/2}$, CV = 1.3, 49923 spikes. {\em A} and
{\em E}: Dependence of the angular velocity $\dot{\theta}$ on the
radial coordinate $r$. {\em B} and {\em F}: Spectra of eigenvalues.
Two almost degenerate outliers are clearly seen. {\em C} and {\em
G}: Eigenvectors in the temporal domain. {\em D} and {\em H}:
Eigenvectors in the frequency domain, plotted in logarithmic scale.}
\end{figure}
This means that the whole reduced system (\ref{ho1})-(\ref{ho2})
revolves with a unique angular velocity $\beta$ around the origin.
In Fig.~\ref{fig6}, the results obtained for two different values of
$\beta$ are shown. In the upper panels, $\beta = 1 /$ ms, in the
lower ones, $\beta = 1/$2 ms. In {\em A} and {\em E}, the dependence
of $\dot{\theta}$ with the radial variable $r$ is shown. In {\em B}
and {\em F}, we see the spectra of eigenvalues. In both cases, two
almost degenerate eigenvalues (labeled {\em a} and {\em b}) are
significantly smaller than unity. The eigenvectors corresponding to
these two eigenvalues are shown in {\em C} and {\em G}. For each
value of $\beta$, the eigenvectors {\em a} and {\em b} are very
similar to each other: they constitute an almost sinusoidal
wave-form, of a very well defined frequency. They only differ in a
$\pi / 2$ phase shift, implying that the relevant stimulus
eigenspace is comprised of all linear combinations of a sine and a
cosine function, of a specific frequency. The frequency, though,
varies with $\beta$. Panels {\em D} and {\em H} show the power
spectra of the two eigenvectors, in logarithmic scale. When $\beta =
1 /$ms, the maximum power is found at a frequency of 1000 Hz ({\em
D}), whereas for $\beta = 0.5$/ms, the eigenvectors peak at $500$ Hz
({\em H}). This means that we are in the presence of a resonance
phenomenon: spiking probability is maximized shortly after stimulus
segments that contain significant power in the characteristic
frequency of the angular motion of the dynamical system.

\begin{figure}[htdf]
\centerline{\includegraphics[keepaspectratio=true, clip = true,
scale = 1.0, angle = 0]{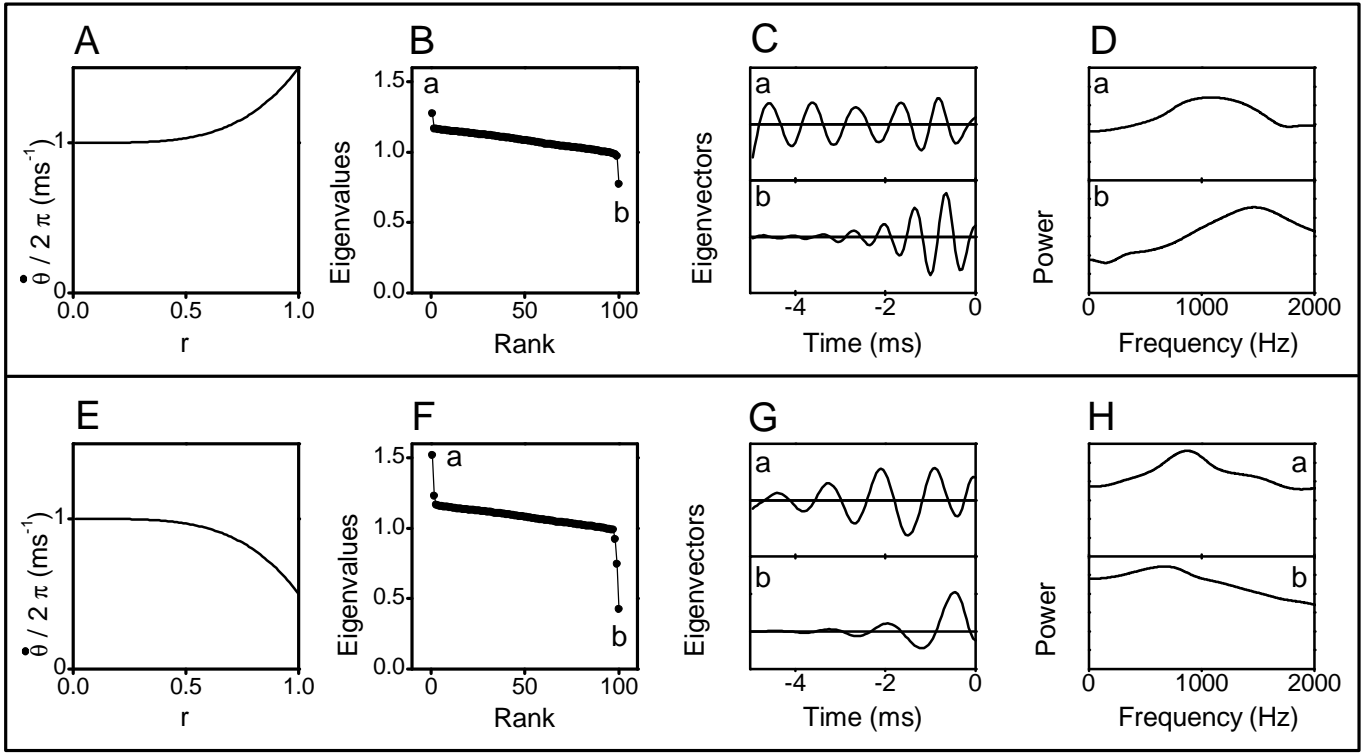}} \caption{\label{fig7} Covariance
analysis of a normal-form type II neuron model in which the angular
velocity $\dot{\theta}$ either increases with the radial coordinate
$r$ (upper panels) or decreases with $r$ (lower panels). The
bifurcation parameter $\alpha$ is set to -0.5/ms, and $\tau = 0.1$
ms. The noise $\sigma$ is chosen so as to obtain a firing rate of
200 Hz. {\sl Upper panels}: $\beta = 1/$ms, $g = 0.5$, $\tau = 0.1$
ms, $\sigma = 0.025$/ms$^{1/2}$, CV = 6.0, 234796 spikes. {\sl Lower
panels}: $\beta = 0.5$ms, $g = -0.5$, $\tau = 0.1$ ms, $\sigma =
0.025$ 0.049/ms$^{1/2}$, CV = 1.3, 200212 spikes. {\em A} and {\em
E}: Dependence of the angular velocity $\dot{\theta}$ on the radial
coordinate $r$. {\em B} and {\em F}: Spectra of eigenvalues. Two
outliers are clearly seen, one of them ({\sl a}) with increased
variance, the other one ({\sl b}) with decreased variance. {\em C}
and {\em G}: Eigenvectors in the temporal domain. {\em D} and {\em
H}: Eigenvectors in the frequency domain, plotted in logarithmic
scale.}
\end{figure}

How do these results change when the angular velocity $\dot{\theta}$
depends on the radial coordinate $r$? This question is pertinent,
since type II systems undergo two bifurcations, in two different
locations in phase space, and are therefore not amenable of a local
reduction. The firing limit cycle appears through a global
bifurcation that in principle, may take place far away from the
resting fixed point. Hence, the frequency associated to the spiking
limit cycle need not coincide with the frequency of subthreshold
oscillations. Therefore, in Fig.~\ref{fig7} we
explore the behavior of two other type II systems, where the angular
velocity $\dot{\theta}$ either increases with $r$ (upper panels) or
decreases with $r$ (lower panels). We see that once the angular
velocity $\dot{\theta}$ sweeps over a whole range of frequencies,
the degeneracy of the relevant eigenvectors is removed. Actually,
now one of the eigenvectors ({\sl a}) is associated to a direction
of increased variance, and the other one ({\sl b}) corresponds to a
diminished variance. Eigenvector {\sl b} is still centered around
the frequency of subthreshold oscillations near the fixed point
(that is, 1000 Hz). However, the dominant frequency of eigenvector
{\sl b} has now shifted to a lower or higher value, depending on
whether $\dot \theta$ is an increasing or decreasing function of
$r$. In the upper panels, $\dot{\theta}$ grows with $r$, implying
that the firing limit cycle has a higher frequency than the
subthreshold oscillations. Correspondingly, the dominant frequency
of eigenvector {\sl b} is shifted to larger values (see {\em D}).
The lower panels, instead, show a case in which $\dot{\theta}$ is a
decreasing function of $r$, and correspondingly, the dominant
frequency of eigenvector {\sl b} has shifted to lower values (see
{\em H}).

\begin{figure}[htdf]
\centerline{\includegraphics[keepaspectratio=true, clip = true,
scale = 1.0, angle = 0]{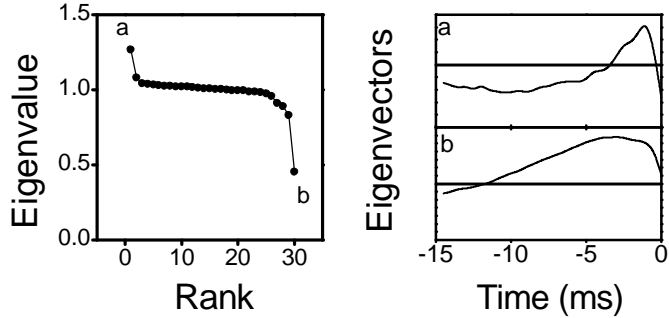}} \caption{\label{fig8} Covariance
analysis of a normal-form type II neuron model in which the
parameters have been chosen in order to obtain two eigenvectors that
span a subspace that is qualitatively similar as the one generated
by the eigenvectors of the HH model (see Fig.~\ref{fig5}). Here,
$\tau = 0.2$ ms, $c= 1.69$/ms, $f = -1.69$/ms, $\beta = 0.0477$/ms,
$d = -0.0183$/ms, $g = 0$, $\alpha = -0.3$/ms, $\sigma =
0.028/\rm{ms}^{1/2}$, CV = 0.79, 49409 spikes. The firing rate was
21.0 Hz.}
\end{figure}

In an arbitrary type II system, the angular velocity may show a
rather complicated dependence on the radial coordinate $r$. In
consequence, the prototypical system described by Eqs.~(\ref{ho1})
and (\ref{ho2}) contains several non-trivial parameters. The shape
of the spectrum of eigenvalues depends rather critically on these
parameters, and so do the associated eigenvectors. Actually, by
choosing those parameters carefully, it is possible to obtain
eigenvalues and eigenvectors that  are similar to those of the
original HH model. In Fig.~(\ref{fig8}) we show the results of
stimulating the normal form Eqs.~(\ref{ho1}) and (\ref{ho2}) with
conveniently selected values of the parameters $\alpha, c, f, \beta,
d, g$ so as to obtain an eigenspace that is qualitatively similar to
the one generated by the eigenvectors of the HH model of
Fig.~\ref{fig5}.
Notice that with these parameters, ${\rm d}\theta / {\rm d}t$ is a
decreasing function of $r$. This is in agreement with the behavior
of the HH model, where the frequency associated to the firing limit
cycle is smaller than the frequency of the subthreshold oscillations
around the resting state.

\section{Discussion}

In this paper, we have analyzed the stimulus features that are
relevant in shaping the spiking probability for type I and type II
neuron models. Type I models undergo a saddle-node, away-limit-cycle
bifurcation at firing onset, and they can be reduced to a normal
form that contains a single scale parameter that determines the
firing frequency. Our analysis shows that when type I models are
stimulated with random input currents whose mean value lies slightly
below threshold, they fire in response to depolarizing stimuli. By
scaling the relevant eigenvectors obtained from a covariance
analysis, we have shown that type I models are not selective to
specific temporal scales: the faster the depolarizing input, the
faster the neural response.

Previous studies (Ag\"uera y Arcas \& Fairhall, 2003) have shown
that the firing probability of integrate-and-fire neurons is
determined by a 2-dimensional space, spanned by the eigenvectors
associated to a small-variance and a large-variance eigenvalues. The
eigenvector associated to the small eigenvalue essentially coincided
with the eigenvector shown in the type I models of this paper of
Sect. \ref{type1}. The second eigenvector obtained for the
integrate-and-fire model, however, combines a hyperpolarizing and a
depolarizing phase. In the models studied in Sect. \ref{type1}, that
eigenvector could sometimes be identified (see the largest
eigenvalue of Fig.~\ref{fig4}). However, as the DC component of the
input current approaches the threshold current, the eigenvalue was
shown to decrease, until it was hidden in the baseline level of all
the other eigenvalues (see Fig.~\ref{fig3}). This effect, however,
is not found in the integrate-and-fire model, for which all input
currents are represented by a spectrum of eigenvalues showing two
outliers. Hence, integrate-and-fire neurons do not behave as our
prototypical type I neurons. The reason for this discrepancy lies in
the fact that near threshold, the slow evolution of
integrate-and-fire neurons revolves around the zone of phase space
representing the firing threshold. In the type I neuron models
discussed in this paper, slow dynamics occurs in the vicinity of the
subthreshold resting state. Hence, the two models cannot be mapped
onto one another, without the introduction of additional abrupt
dynamical features, as the voltage cutoffs employed in Hansel \&
Mato (2003).

In contrast, type II models are governed by a subcritical Hopf
bifurcation, thereby requiring a 2-dimensional, non local
description. As a consequence, type II models are characterized by
several parameters, that retain various temporal properties of the
original dynamical system. These differences between type I and type
II neuron models imply that each one of them is responsive to
particular features of the input current. Baroni and Varona (2007)
have suggested that the differences in the input/output
transformation carried out by type I and type II neurons may have
evolved to activate different populations of neurons, depending on
the specific temporal patterns of the presynaptic input. Muresan and
Savin (2007), in turn, have discussed the consequences of these
differences at the network level. While type II neurons favor
self-sustainability of network activity, type I cells increase the
richness of the responsiveness to external stimuli.

The main distinction between the stimulus features driving type I
and type II models, as demonstrated by our covariance analysis,
provides further justification to the distinction introduced by
Izhikevich (2001), where type I neurons are called {\em
integrators}, and Hopf-like type II neurons are {\em resonators}.
Our work shows that, indeed, near threshold type I neurons are
essentially waiting to receive excitatory input (no matter their
time scale), whereas Hopf-like type II neurons are able to detect
input segments that contain sufficient power in the frequency band
that is well represented by their intrinsic dynamics.

Can the relevant stimulus features be connected to the
phase-resetting curves? By comparing  Fig. \ref{fig2} with Fig.
\ref{fig3}, we conclude that both the phase-response curve and the
preferred stimulus feature of type I neurons (which, as stated
above, coincides with the STA) are monophasic and positive. In turn,
both the eigenvector corresponding to the smallest eigenvalue of
type II neurons (the one that is most similar to the STA) and the
phase-response curve exhibit positive and negative phases (compare
Figs. \ref{fig2} and \ref{fig5}). This similarity, however, can only
be stated at a qualitative level. Recall that phase-response curves
can only be defined in the supra-threshold regime, whereas our the
covariance analysis concerned random input currents of subthreshold
mean.

\begin{figure}[htdf]
\centerline{\includegraphics[keepaspectratio=true, clip = true,
scale = 1.0, angle = 0]{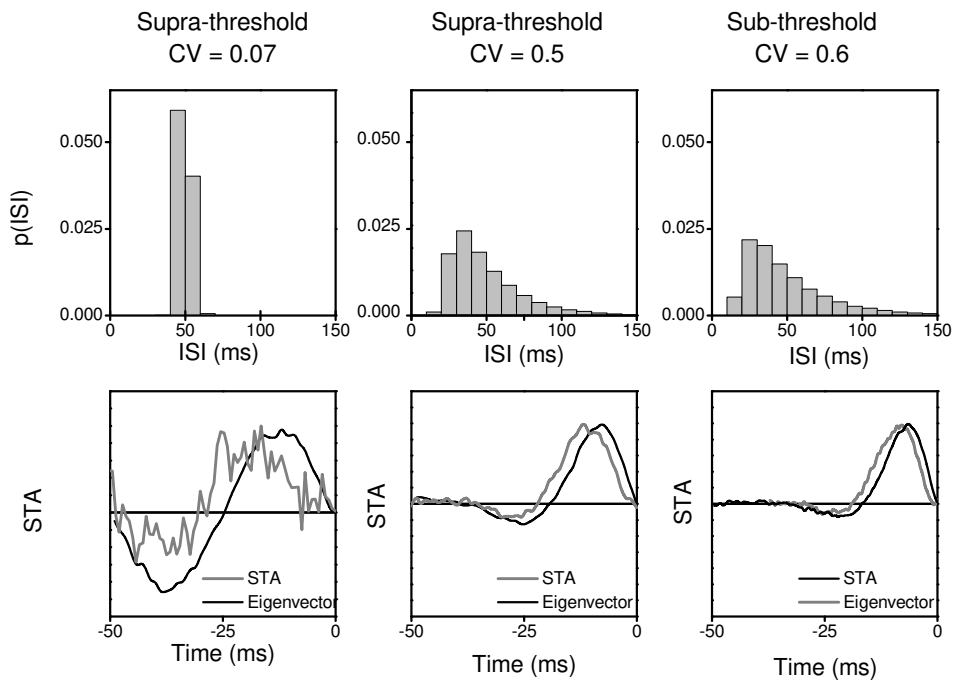}} \caption{\label{fig9} ISI
probability densities (upper panels) and most relevant stimulus
feature (lower panels) of a normal-form, type I neuron firing at 20
Hz, for three different stimulation protocols. The most relevant
stimulus feature is obtained with two different methods: the STA
(black line) and the most prominent eigenvector (grey line). {\em
Left}: Regular firing, with stimuli of supra-threshold mean. $\alpha
= 0.004$, $\sigma = 0.00029$ ms$^{1/2}$, $\tau = 0.2$ ms, 500159
spikes, CV = 0.07. {\em Middle}: Irregular firing, with stimuli of
small supra-threshold mean. $\alpha = 0.002$, $\sigma = 0.00295$
ms$^{1/2}$, $\tau = 0.2$ ms, 201287 spikes, CV = 0.5. {\em Right}:
Irregular firing, with stimuli of small sub-threshold mean. $\alpha
= -0.001$, $\sigma = 0.004345$ ms$^{1/2}$, $\tau = 0.2$ ms, 49462
spikes, CV = 0.6. When the system is firing regularly (low CV), the
STA shows a positive and a negative phase, resembling the derivative
of the phase-resetting curve. In contrast, when the system fires in
an irregular fashion (large CV), the STA exhibits a large positive
phase, and a small negative portion.}
\end{figure}

Ermentrout, Gal\'an, \& Urban (2007) have proved that when a cell is
firing regularly while receiving mild random perturbative currents,
its spike-triggered average is equal to the temporal derivative of
the phase-resetting curve. This result contrasts with our
observation that the phase-response curve itself (and not its
derivative) has the same sign as the relevant stimulus features
inducing spiking, for subthreshold stimuli. The two conclusions,
however, are not in conflict with one another. In Fig. \ref{fig9},
the ISI distribution, the STA, and the most relevant eigenvector of
a normal-form, type I neuron is shown, when driven with three
different stimulation protocols.
The plots on the left correspond to an input stimuli whose mean
value is sufficiently large as to keep the system firing regularly
at 20 Hz. The noisy component of the injected current, therefore,
only rarely modifies the regular spiking pattern, occasionally
advancing or delaying an action potential. As a consequence, the
inter-spike-interval (ISI) distribution reduces to a very narrow
peak located at 50 ms. The STA, in turn, is clearly bi-phasic, and
constitutes a very good approximation of the temporal derivative of
the phase-resetting curve (see Ermentrout, Gal\'an and Urban, 2007).
The most relevant eigenvector is qualitatively similar to the STA,
though its shape is noticeably more sensitive to limited sampling.

In order to explore how these results evolve as the activity of the
cell becomes more irregular, in the middle and right panels we show
the results obtained with other stimulating scheme, where the mean
stimulus is lowered, and the stochastic component of the input
current is increased as to maintain a 20 Hz firing rate. The ISI
distributions are therefore significantly widened. We see that the
STA is no longer symmetric with respect to the mean stimulus: the
depolarized phase of the STA is markedly larger than the
hyperpolarized one. As the mean stimulus is made increasingly
negative, the negative phase disappears completely, and the STA
merges into the curves shown in Fig.~\ref{fig4}. We therefore
conclude that the exact relationship between the relevant stimulus
features inducing spiking and the phase-response curve depends
critically on how regular the spike train is. Which, for fixed
firing rate, is governed by the ratio between the SD of the stimulus
and its mean value (the CV). We hope that the present analysis may
serve to inspire future investigations in the connection between
these quantities.

\section{Acknowledgements}

We thank Eugenio Urdapilleta for useful discussions. This work has
been supported by the Comisi\'on Nacional de Investigaciones
Cient\'{\i}ficas y T\'ecnicas (PIP 5140), the Alexander von Humboldt
Foundation, the Universidad Nacional de Cuyo and the Agencia
Nacional de Promoci\'on Cient\'{\i}fica y Tecnol\'ogica.

\newpage

\section*{Appendix A: Wang Buzsaki (WB) Model}

The dynamical equations for the conductance-based type I neuron
model used in this work were introduced by Wang and Buzsaki (1996).
They are
\begin{eqnarray}
\label{WBeq} C \frac{{\rm d} V}{{\rm d}t} &=& I -g_{\rm Na} {\rm
m}_{\infty}^3(V)  h (V- V_{\rm Na}) - g_{\rm K} n^4 (V - V_{\rm K})
- g_{\ell} (V-V_\ell)
\label{wb1}  \\
\frac{{\rm d}h}{{\rm d}t} &=& \frac{{\rm h}_{\infty}(V) - h}{\tau_{\rm h}(V)} \\
\frac{dn}{dt} &=& \frac{{\rm n}_{\infty} (V) - n}{\tau_{\rm n}(V)}.
\end{eqnarray}
The parameters $g_{\rm Na}$, $g_{\rm K}$ and $g_\ell$ are the
maximum conductances per surface unit for the sodium, potassium and
leak currents respectively and  $V_{\rm Na}$, $V_{\rm K}$ and
$V_\ell$ are the corresponding reversal potentials. The capacitance
per surface unit is denoted by $C$. The external stimulus on the
neuron is represented by an external current $I$. The functions
${\rm m}_{\infty}(V)$, ${\rm h}_{\infty}(V)$, and ${\rm
n}_{\infty}(V)$ are defined as ${\rm x}_{\infty}(V) = a_{\rm
x}(V)/[a_{\rm x} (V) + b_{\rm x}(V)]$, where ${\rm x} = {\rm m},
{\rm n}$ or ${\rm h}$. In turn, the characteristic times (in
milliseconds) $\tau_{\rm n}$ and $\tau_{\rm h}$ are given by
$\tau_{\rm x} = 1/[a_{\rm x}(V) + b_{\rm x}(V)]$, and
\begin{eqnarray}
a_{\rm m} &=& -0.1(V + 35) / (\exp(-0.1 (V + 35))-1),\\
b_{\rm m} &=& 4 \exp(-(V + 60) / 18),\\
a_{\rm h} &=& \phi \, 0.07 \exp(-(V + 58) / 20),\\
b_{\rm h} &=& \phi/(\exp(-0.1(V + 28))+1).
\end{eqnarray}
The other parameters of the sodium current are: $g_{\rm Na} = 35
\rm{mS/cm}^2$ and $V_{\rm Na} = 55 \rm{mV}$. The delayed rectifier
current is described in a similar way as in the HH model with:
\begin{eqnarray}
a_n &=& \phi \, 0.01(V + 34)/(1-\exp(-0.1(V + 34))),\\
b_n &=& \phi \, 0.125 \exp(-(V + 44) / 80).
\end{eqnarray}
Other parameters of the model are: $V_{\rm K} = -90 \rm{mV}, V_{\rm
Na}=55 \rm{mV}, V_\ell=-65 \rm{mV}, C=1 \mu \rm{F}/\rm{cm}^2$,
$g_\ell=0.1 \rm{mS/cm}^2, g_{\rm Na}= 35 \rm{mS/cm}^2, g_{\rm K} = 9
\rm{mS/cm}^2, \phi=3$.

\section*{Appendix B: the Hodgkin-Huxley (HH) model}

The dynamical equations of the Hodgkin-Huxley (Hodgkin \& Huxley,
1952) model read
\begin{eqnarray}
\label{HHeq} C \frac{{\rm d}V}{{\rm d}t} &=& I -g_{\rm Na} m^3 h
(V- V_{\rm Na}) - g_{\rm K} n^4 (V-V_{\rm K}) - g_\ell (V-V_\ell),  \\
\frac{{\rm d}m}{{\rm d}t} &=& \frac{{\rm m}_{\infty}(V) -
m}{\tau_{\rm m}(V)},
\label{hh1} \\
\frac{{\rm d}h}{{\rm d}t} &=& \frac{{\rm h}_{\infty}(V) - h}{\tau_{\rm h}(V)}, \\
\frac{{\rm d}n}{{\rm d}t} &=& \frac{{\rm n}_{\infty} (V) -
n}{\tau_{\rm n}(V)}.
\end{eqnarray}
For the squid giant axon, the parameters at $6.3 \ ^0C$ are: $V_{\rm
Na} = 50 \rm{mV}$, $V_{\rm K} = -77$ \rm{mV}, $V_\ell=-54.4
\rm{mV}$, $ g_{\rm Na} = 120 \rm{mS/cm}^2$, $g_{\rm K} = 36
\rm{mS/cm}^2$ , $g_\ell =0.3 \rm{mS/cm}^2$, and $C = 1 \mu
\rm{F}/\rm{cm}^2$. The functions ${\rm m}_{\infty}(V)$, ${\rm
h}_{\infty}(V)$, ${\rm n}_{\infty}(V)$, $\tau_{\rm m}(V)$,
$\tau_{\rm n}(V)$, and $\tau_{\rm h}(V)$, are defined in terms of
the functions $a(V)$ and $b(V)$ as in the WB model, but now
\begin{eqnarray}
a_{\rm m} &=& 0.1 (V + 40)/(1-\exp ((-V - 40) / 10)),\\
b_{\rm m} &=& 4 \exp ((-V - 65) / 18),\\
a_{\rm h} &=& 0.07 \exp ((-V - 65) / 20)),\\
b_{\rm h} &=& 1/(1+\exp ((-V - 35) / 10),\\
a_{\rm n} &=& 0.01 (V + 55)/(1 - \exp ((-V - 55) / 10)),\\
b_{\rm n} &=& 0.125 \exp ((-V - 65) /80).
\end{eqnarray}

\newpage

\begin{center}
{\bf References}
\end{center}

\begin{enumerate}

\item[-] Ag\"uera y Arcas, B., \& Fairhall, A. L. (2003). What causes a
neuron to spike? {\em Neural Comput. 15} 1789-1807.

\item[-] Ag\"uera y Arcas, B., Fairhall, A. L., \& Bialek W. (2003).
Computation in Single Neuron: Hodgkin and Huxley Revisited. {\em
Neural Comput. 15}  1715-1749.

\item[-] Baroni F. \& Varona P. (2007). Subthreshold oscillations
and neuronal input-output relationships. {\em Neurocomput. 70}:
161-1614.

\item[-] Bialek W. \& De Ruyter von Steveninck R. R. (2003). Features
and dimensions: motion estimation in fly vision. q-bio/0505003.

\item[-] Brown E., Moehlis J., \& Holmes P. (2004) On the Phase
Reduction and Response Dynamics of Neural Oscillator Populations.
{\em Neural Comput. 16} 673-715.

\item[-] Ermentrout, B. (1996) Type I Membranes, Phase Resetting Curves,
and Synchrony. {\em Neural Comp. 8} 979-1001.

\item [-] Ermentrout, B. \& Kopell, N. (1986) Parabolic Bursting in an Excitable System
Coupled with a Slow Oscillation {\em SIAM Journal on Applied Mathematics 46} 233-253.

\item [-] Ermentrout, B., Gal\'an, R., \& Urban, N. (2007) Relating neural
dynamics to neural coding, arXiv0707:0245v2 [q-bio.NC].

\item[-] Fairhall A., Burlingame C. A., Narasimhan R., Harris R. A.,
Puchalla J. L. \& Berry M. J. (2006). Selectivity for Multiple Stimulus Features
in Retinal Ganglion Cells. {\em J. Neurophysiol. 96} 2724-2738.

\item[-] Hansel D., Mato G., \& Meunier C. (1995). Synchrony in
excitatory neural networks. {\em Neural Comput 7} 307-337 (1995).

\item[-] Hansel D. \& Mato G. (2003) Asynchronous states and the emergence of synchrony
in large networks of interacting excitatory and inhibitory neurons.
{\em Neural Comput 15} 1-56.

\item[-] Hodgkin A. L. (1948). The local electric charges associated
with repetitive action in non-medulated axons. {\em J. Physiol. 107}
165-181.

\item[-] Hodgkin A. L. \& Huxley A. F. (1952) A quantitative
description of membrane current and application to conductance in
excitation nerve. {\em J. Physiol. (London) 117} 500-544.

\item[-] Hong S., Ag\"uera y Arcas B., \& Fairhall A. (2007).
Single neuron computation: from dynamical system to feature
detector. arXiv:q-bio/0612025

\item[-]Izhikevich, E. (2001). Resonate and fire neurons. {\em Neural
Networks 14} 833-894.

\item[-]Izhikevich, E. (2007). {\em Dynamical Systems in Neuroscience.
The Geometry of Excitability and Bursting}. The MIT Press.

\item[-]Kuramoto, Y. (2003). {\em Chemical Oscillations, Waves and
Turbulence.} New York: Dover.

\item[-] Maravall M., Petersen R. S., Fairhall A.L., Arabzadeh E. \&
Diamond M. E. (2007). Shifts in Coding Properties and Maintenance of
Information Transmission during Adaptation in Barrel Cortex. {\em
PLoS Biol. 5(2)} e19.

\item[-] Muresan R.C. \& Savin C. (2007). Resonance or Integration?
Self-sustained Dynamics and Excitability of Neural Microcircuits.
{\em J. Neurophysiol.} 97: 1911-1930.

\item[-] Paninski, L. (2003). Convergence properties of some spike-triggered
analysis techniques. {\em Network 14} 437-464.


\item[-]  Rust N. C., Schwartz O., Movshon J. A, \& Simoncelli E. P.
(2005) Spatiotemporal Elements of Macaque V1 Receptive Fields. {\em
Neuron 46} 945-956.

\item[-] Schwartz O., Pillow J. W., Rust N. C. \& Simoncelli E. P.
(2006). Spike-triggered neural characterization. {\em J. Vision 6}
484-507.

\item[-] Wang, X. J., \& Buzs\'aki G (1996) Gamma Oscillation by Synaptic
Inhibition in a Hippocampal Interneuronal Network Model. {\em J.
Neurosci. 16} 6402-6413.
\end{enumerate}

\end{document}